
%
\input phyzzx
\catcode`\@=11
\newfam\mibfam
%
\font\sevensl =cmsl8 at 7pt
\font\sevenit =cmti7
\font\seventeenmib =cmmib10 scaled\magstep3 \skewchar\seventeenmib='177
\font\fourteenmib =cmmib10 scaled\magstep2 \skewchar\fourteenmib='177
\font\twelvemib =cmmib10 scaled\magstep1 \skewchar\twelvemib='177
\font\tenmib =cmmib10 \skewchar\tenmib='177
\font\ninemib =cmmib9 \skewchar\ninemib='177
\font\sixmib =cmmib6 \skewchar\sixmib='177
%
\def\fourteenf@nts{\relax
\textfont0=\fourteenrm \scriptfont0=\tenrm
\scriptscriptfont0=\sevenrm
\textfont1=\fourteeni \scriptfont1=\teni
\scriptscriptfont1=\seveni
\textfont2=\fourteensy \scriptfont2=\tensy
\scriptscriptfont2=\sevensy
\textfont3=\fourteenex \scriptfont3=\twelveex
\scriptscriptfont3=\tenex
\textfont\itfam=\fourteenit \scriptfont\itfam=\tenit
\textfont\slfam=\fourteensl \scriptfont\slfam=\tensl
\textfont\bffam=\fourteenbf \scriptfont\bffam=\tenbf
\scriptscriptfont\bffam=\sevenbf
\textfont\mibfam=\fourteenmib \scriptfont\mibfam=\tenmib
\textfont\ttfam=\fourteentt
\textfont\cpfam=\fourteencp }
\def\twelvef@nts{\relax
\textfont0=\twelverm \scriptfont0=\ninerm
\scriptscriptfont0=\sixrm
\textfont1=\twelvei \scriptfont1=\ninei
\scriptscriptfont1=\sixi
\textfont2=\twelvesy \scriptfont2=\ninesy
\scriptscriptfont2=\sixsy
\textfont3=\twelveex \scriptfont3=\tenex
\scriptscriptfont3=\tenex
\textfont\itfam=\twelveit \scriptfont\itfam=\nineit
\textfont\slfam=\twelvesl \scriptfont\slfam=\ninesl
\textfont\bffam=\twelvebf \scriptfont\bffam=\ninebf
\scriptscriptfont\bffam=\sixbf
\textfont\mibfam=\twelvemib \scriptfont\mibfam=\ninemib
\scriptscriptfont\mibfam=\sixmib
\textfont\ttfam=\twelvett
\textfont\cpfam=\twelvecp }
\def\tenf@nts{\relax
\textfont0=\tenrm \scriptfont0=\sevenrm
\scriptscriptfont0=\fiverm
\textfont1=\teni \scriptfont1=\seveni
\scriptscriptfont1=\fivei
\textfont2=\tensy \scriptfont2=\sevensy
\scriptscriptfont2=\fivesy
\textfont3=\tenex \scriptfont3=\tenex
\scriptscriptfont3=\tenex
\textfont\itfam=\tenit \scriptfont\itfam=\sevenit
\textfont\slfam=\tensl \scriptfont\slfam=\sevensl
\textfont\bffam=\tenbf \scriptfont\bffam=\sevenbf
\scriptscriptfont\bffam=\fivebf
\textfont\mibfam=\tenmib
\textfont\ttfam=\tentt
\textfont\cpfam=\tencp }
\def\mib{\n@expand\f@m\mibfam}
\Twelvepoint
\catcode`\@=12
\tolerance=9999
\overfullrule=0pt
\def\e{{\rm e}}
\def\del{\partial}
\def\dslash{\del\kern-0.55em\raise 0.14ex\hbox{/}}
\def\Lag{{\cal L}}

\def\bfx{{\mib x}}
\def\bfpT{{{\mib p}_T}}
\def\abs#1{{\left|{#1}\right|}}
\def\expecv#1{\langle #1 \rangle}
\def\phipa{\phi^{(+\alpha)}}
\def\phima{\phi^{(-\alpha)}}

\def\a{\alpha}
\def\b{\beta}
\def\ep{\epsilon}
\def\g{\gamma}

\def\s{\sigma}
\def\t{\tau}
%
%
\def\papersize{\hsize=37pc \vsize=52pc \hoffset=1.5mm \voffset=1pc
\advance\hoffset by\HOFFSET \advance\voffset by\VOFFSET
\pagebottomfiller=0pc
\skip\footins=\bigskipamount \normalspace }
\papersize
\date={23rd May, 1994}
\Pubnum={}
\Pubtype={}
\titlepage
\centerline{}
\vskip -1.3cm
\leftline{ICHEP94 Ref. gls0313}
\leftline{Submitted to Pa 12, Pl 18}
\title{\fourteenbf Fermion Scattering off Electroweak Bubble Wall
 and Baryogenesis}
\bigskip\bigskip
\centerline{Koichi Funakubo${}^{1)}$\footnote{\#a}
{e-mail: \ funakubo@sagagw.cc.saga-u.ac.jp},\ \ \  Akira Kakuto${}^{2)}$%
\footnote{\#b}{e-mail: \ kakuto@fuk.kindai.ac.jp},
\ \ Shoichiro Otsuki${}^{2)}$,}
\centerline{Kazunori Takenaga${}^{3)}$\footnote{\#c}%
{e-mail: \ take1scp@mbox.nc.kyushu-u.ac.jp}
\ \ and\ \ \  Fumihiko Toyoda${}^{2)}$\footnote{\#d}%
{e-mail: \ ftoyoda@fuk.kindai.ac.jp}}
\bigskip
\centerline{${}^{1)}$\it Department of Physics, Saga University,
Saga 840 Japan}
\centerline{${}^{2)}$\it Department of Liberal Arts, Kinki University in
Kyushu, Iizuka 820 Japan}
\centerline{${}^{3)}$\it Department of Physics, Kyushu University,
Fukuoka 812 Japan}
\bigskip\bigskip
\abstract{
By treating CP-violating interaction as a perturbative term, 
we solve the Dirac equation in the background of electroweak bubble wall
(the distorted wave Born approximation). We obtain the transmission and
reflection coefficients for a chiral fermion incident from the
symmetric-phase region and for the one from the broken-phase region
respectively. There hold the respective sets of unitarity relations and
also reciprocity relations among them. These relations enable us to rigorously
derive quantum-number flux through the bubble wall, which is the first order
quantity of the CP violation. The flux is found to be negligible for a
thick wall such that $m_0/a\gsim 2$, where $1/a$ is the wall thickness
and $m_0$ is the fermion mass.}
\vfill\eject
\chapter{Introduction}
It is well known that electroweak theory satisfies the three necessary
conditions by Sakharov${}^{1)}$ to generate the baryon asymmetry of the
universe, provided that the phase transition is first order${}^{2)}$.
During the phase-transition process of first order, we assign
a complex mass to a fermion as a function of $z$, $m(z) = m_R(z) + im_I(z)$,
where $z$ is the coordinate perpendicular to the wall.
The real part $m_R(z)$ asymptotically behaves such that
$m_R(z)\rightarrow 0$ as $z\rightarrow -\infty$ (symmetric phase)
and $m_R(z)\rightarrow m_0$ as $z\rightarrow +\infty$ (broken phase),
where $m_0$ is the fermion mass. The imaginary part $m_I(z)$
produces quantum-number flow through the bubble wall${}^{3)}$, 
if $m_I(z) / m_R(z)$ is not a constant as in the case that CP is
violated in the Higgs sector.\par
We give a general prescription to treat fermion propagation
in CP-violating bubble-wall background, by regarding the CP-violating term as
a small perturbation (DWBA---the distorted wave Born approximation)${}^{4)}$.
We obtain the transmission and reflection coefficients for chiral fermions
incident from the symmetric-phase or from the broken-phase region.
There hold the respective sets of unitarity relations and also reciprocity
relations among the coefficients. These relations enable us to rigorously
derive quantum-number flux through the bubble wall, which is the
first order quantity of the CP violation. Here the dynamical quantity,
$\Delta R\equiv R^s_{R\rightarrow L} - \bar R^s_{R\rightarrow L}$,
where $R^s_{R\rightarrow L}$ ($\bar R^s_{R\rightarrow L}$)
is the reflection coefficient for right-handed chiral fermion
(anti-fermion) incident from the symmetric-phase region,
is of primary importance. If $\abs {\Delta R}$ is small, no sufficient
amount of baryon-number asymmetry of the universe is generated${}^{4,5)}$.
\par
Taking $m_R(z)$ of the kink type${}^{6)}$, we evaluate $\Delta R$ for several
forms of $m_I(z)$. $\abs {\Delta R}$ is extremely small for a thick wall
such that the thickness $1/a$ is larger than $2/m_0$. For one-Higgs-doublet
models with $1/a\sim(20-40)/T$ ${}^{7)}$, this excludes the top quark from
the baryogenesis game.
%
%
\chapter{DWBA to CP-Violating Dirac Equation}
\noindent
\section{Dirac equation and Ansatz}
We consider one-flavor model described by the lagrangian,
$$
  \Lag = \bar\psi_L i\dslash \psi_L + \bar\psi_R i\dslash \psi_R
         + ( f \bar\psi_L\psi_R\phi + {\rm h.c.} ).
\eqn\IIa
$$
In the vacuum, near the first-order phase transition point,
$\expecv\phi$ may be $x$-dependent field, so that we put
$$
  m(\bfx) = - f \expecv\phi(\bfx),
\eqn\IIb
$$
where $m(\bfx)$ is complex-valued and we neglect the time dependence. If the
phase of $m(\bfx)$ has no spatial dependence, it is removed by a constant
bi-unitary transformation, which is outside of our interest.
The Dirac equation describing fermion propagation in the bubble-wall
background  with CP violation is${}^{3)}$
$$
i\dslash\psi(t,\bfx)-m(\bfx)P_R\psi(t,\bfx)-m^*(\bfx)P_L\psi(t,\bfx) = 0.
\eqn\IIc
$$
For the bubble wall with large enough radius, $m(\bfx)$ could be regarded as
a function of only one spatial coordinate, so that we put
$m(\bfx)=m(z)$.\par
To solve \IIc, we take the following Ansatz${}^{6)}$:
$$\eqalign{
 \psi(t,\bfx)
 &= (i\dslash + m^*(z)P_R + m(z)P_L)
    \e^{i\s(-Et+\bfpT\bfx_T)}\psi_E(\bfpT,z) \cr
 &= \e^{i\s(-Et+\bfpT\bfx_T)}[\s(\g^0E-\g_Tp_T)
    + i\g^3\del_z+ m^*(z)P_R + m(z)P_L] \psi_E(\bfpT,z), \cr
}\eqn\IId
$$
where $\s=+(-)$ for positive (negative)-energy solution,
$\bfpT=(p^1,p^2)$, $\bfx_T=(x^1,x^2)$, $p_T=\abs{\bfpT}$ and
$\g_Tp_T=\g^1p^1+\g^2p^2$. By putting $E=E^*\cosh\eta$ and
$p_T=E^*\sinh\eta$ with $E^*=\sqrt{E^2-p_T^2}$, the Lorentz
transformation eliminates $\bfpT$.
After this Lorentz rotation for a fixed $\bfpT$,
the Dirac equation is rewritten as
$$
 \bigl[ {E^*}^2 + \del_z^2-{\abs{m(z)}}^2
    +im_R^\prime(z)\g^3 - m_I^\prime(z)\g_5\g^3 \bigr]\psi_E(z)=0,
\eqn\IIe
$$
where $m(z)=m_R(z)+im_I(z)$. Now let us introduce a set of dimensionless
variables using a parameter $a$, whose inverse characterizes the thickness
of the wall: $m_R(z)=m_0f(az)=m_0f(x)$, $m_I(z)=m_0g(az)=m_0g(x)$,
$x\equiv az$, $\t\equiv at$, $\ep \equiv E^*/a$, $\xi\equiv m_0/a$,
where $m_0$ is the fermion mass in the broken phase. Eq.\IIe\ is expressed as
$$
 [\ep^2 + {d^2\over{dx^2}} - \xi^2(f(x)^2+g(x)^2)
  +i\xi f^\prime(x)\g^3-\xi g^\prime(x)\g_5\g^3]
  \psi_\ep(x) = 0.
\eqn\IIh
$$
\par
As for $f(x)$ and $g(x)$, we do not specify their functional
forms but only assume that
$$
 f(x)\rightarrow\cases{1, &as $x\rightarrow+\infty$,\cr
                       0, &as $x\rightarrow-\infty$,\cr}
\eqn\IIi
$$
and that $\abs{g(x)}<<1$, ${\it {i.e.,}}$ small CP violation.
Eq.\IIi\  means that the system is in the broken (symmetric) phase at
$x\sim+\infty$ ($x\sim-\infty$), the wall height being $m_0$.
\par
\section{DWBA to the Dirac equation${}^{4)}$}
We regard the small $\abs{g(x)}$ as a perturbation, and keep quantities
up to $O(g^1)$. Put
$$
   \psi_\ep(x) = \psi^{(0)}(x) + \psi^{(1)}(x),
\eqn\IIj
$$
where $\psi^{(0)}(x)$ is a solution to the unperturbed equation
$$
 [\ep^2 + {d^2\over{dx^2}} - \xi^2f(x)^2+i\xi f^\prime(x)\g^3]
  \psi^{(0)}(x) = 0
\eqn\IIk
$$
with an appropriate boundary condition. Then $\psi^{(1)}(x)$ of $O(g^1)$
is solved as
$$
 \psi^{(1)}(x) = \int dx^\prime G(x,x^\prime)V(x^\prime)
                 \psi^{(0)}(x^\prime)\qquad {\rm with}\qquad
   V(x) = -\xi g^\prime(x)\g_5\g^3.
\eqn\IIl
$$
$G(x,x^\prime)$ is the Green's function for the operator in \IIk\ satisfying
the same boundary condition as $\psi^{(0)}(x)$. To this order, the solution
to the Dirac equation is given by
$$
 \psi(x)\simeq \e^{-i\s\ep\t}\Bigl\{ [\s\ep\g^0 +
    i\g^3{d\over{dx}}+\xi f(x)][\psi^{(0)}(x) + \psi^{(1)}(x)]
         - i\xi g(x)\g_5\psi^{(0)}(x) \Bigr\}.
\eqn\IIo
$$
If we expand $\psi^{(0)}(x)$ in terms of the eigenspinors of $\g^3$
as $\psi^{(0)}(x)\sim \phi_\pm(x)u^s_\pm$ with
$ \gamma^3 u^s_\pm = \pm i u^s_\pm (s=1,2)$, $\phi_\pm(x)$ satisfies
$$
  [\ep^2 + {d^2\over{dx^2}} - \xi^2f(x)^2 \mp \xi f^\prime(x)]
  \phi_\pm(x) = 0.
\eqn\IIq
$$
Because of \IIi, the asymptotic forms of $\phi_\pm(x)$ should be
$ \phi_\pm(x)\rightarrow \e^{\a x}, \e^{-\a x} (x\rightarrow+\infty)$ and
$\e^{\b x}, \e^{-\b x} (x\rightarrow-\infty),$
where $\a=i\sqrt{\ep^2-\xi^2}$ and $\b=i\ep$. Putting all these together,
we obtain the asymptotic forms of the wave function (2.11) at
$x \rightarrow \pm \infty$.
\section{Fermion incident from symmetric-phase region${}^{4)}$}
We consider a state in which the incident wave coming from $x=-\infty$ is
reflected in part at the bubble wall, while at $x=+\infty$ only the
transmitted wave exists. We denote two independent solutions to \IIq\
as $\phipa_\pm(x)$ and $\phima_\pm(x)=\bigl(\phipa_\pm(x)\bigr)^*$ which
behave as
$$\eqalign{
 \phipa_\pm(x)\rightarrow\e^{\a x},\qquad
 \phima_\pm(x)\rightarrow\e^{-\a x}
}\eqn\IIra
$$
at $x\rightarrow+\infty$.
Their asymptotic forms at $x\rightarrow-\infty$ are
$$\eqalign{
 &\phipa_\pm(x)\sim \g_\pm(\a,\b)\e^{\b x}+\g_\pm(\a,-\b)\e^{-\b x}, \cr
 &\phima_\pm(x)\sim \g_\pm(-\a,\b)\e^{\b x}+\g_\pm(-\a,-\b)\e^{-\b x}.\cr
}\eqn\IIr
$$
{}From these, the general solution to \IIk\ is eventually given as
$$
 \psi^{(0)}(x) = \sum_s
 [A_s^{(-)}\phima_+(x) + A_s^{(+)}\phipa_+(x)]u^s_+.
\eqn\IIv
$$
The required boundary condition is achieved by setting $A_s^{(-)}=0$
for $\s=+$ and $A_s^{(+)}=0$ for $\s=-$ respectively.
The Green's function which matches this boundary condition can be
constructed from $\phi^{(\pm\a)}_\pm(x)$.
\par
{}From the asymptotic forms of
$\bigl(\psi_{\s}(x)\bigr)^{inc}$,
$\bigl(\psi_{\s}(x)\bigr)^{trans}$ and
$\bigl(\psi_{\s}(x)\bigr)^{refl}$ in \IIo,
we obtain those of the vector and axial-vector currents,
$j_V^\mu = \bar\psi\g^\mu\psi$ and $j_A^\mu = \bar\psi\g^\mu\g_5\psi$.
In terms of the chiral currents, $j^\mu_L = (1/2)(j^\mu_V-j^\mu_A)$ and
$j^\mu_R = (1/2)(j^\mu_V+j^\mu_A)$,
the transmission and reflection coefficients for the chiral fermion
are defined as
$$\eqalign{
 &T^{(\s)}_{L\rightarrow L(R)}
 = \bigl(j^3_{L(R),\s}\bigr)^{trans}\bigr|_{A_1^\s=0}
  /\bigl(j^3_{L,\s}\bigr)^{inc},        \cr
 &T^{(\s)}_{R\rightarrow L(R)}
 = \bigl(j^3_{L(R),\s}\bigr)^{trans}\bigr|_{A_2^\s=0}
  /\bigl(j^3_{R,\s}\bigr)^{inc},         \cr
 &R^{(\s)}_{L(R)\rightarrow R(L)}
 = -\bigl(j^3_{R(L),\s}\bigr)^{refl}
   /\bigl(j^3_{L(R),\s}\bigr)^{inc}.        \cr
}\eqn\IIae
$$
If we denote $R^s(T^s)=R^{(+)}(T^{(+)})$ and
$\bar R^s(\bar T^s)=R^{(-)}(T^{(-)})$,
where the superscript $s$ denotes the fermion incident from the
symmetric-phase region, we have
$$\eqalign{
 &T_{L\rightarrow L}^s=\bar T_{R\rightarrow R}^s
={{\sqrt{\ep^2-\xi^2}+\ep}\over{2\ep\abs{\g_+(\a,\b)}^2}}
 (1-\delta^{inc}),     \cr
 &T_{L\rightarrow R}^s=\bar T_{R\rightarrow L}^s
={{\sqrt{\ep^2-\xi^2}-\ep}\over{2\ep\abs{\g_+(\a,\b)}^2}}
 (1-\delta^{inc}),     \cr
 &T_{R\rightarrow L}^s=\bar T_{L\rightarrow R}^s
={{\sqrt{\ep^2-\xi^2}-\ep}\over{2\ep\abs{\g_+(\a,\b)}^2}}
 (1+\delta^{inc}),     \cr
 &T_{R\rightarrow R}^s=\bar T_{L\rightarrow L}^s
={{\sqrt{\ep^2-\xi^2}+\ep}\over{2\ep\abs{\g_+(\a,\b)}^2}}
 (1+\delta^{inc}),     \cr
 &R_{L\rightarrow R}^s=\bar R_{R\rightarrow L}^s
=\abs{\g_+(\a,-\b)\over\g_+(\a,\b)}^2
 (1-\delta^{inc}-\delta^{refl}),   \cr
 &R_{R\rightarrow L}^s=\bar R_{L\rightarrow R}^s
=\abs{\g_+(\a,-\b)\over\g_+(\a,\b)}^2
 (1+\delta^{inc}+\delta^{refl}).    \cr
}\eqn\IIaf
$$
Here the corrections by the CP violation are
$$\eqalign{
 \delta^{inc} &= {\xi\over{2\sqrt{\ep^2-\xi^2}}}
 \Bigl( {\g_-(-\a,\b)\over\g_+(\a,\b)}I + c.c. \Bigr),    \cr
 \delta^{refl} &= {\xi\over{2\sqrt{\ep^2-\xi^2}}}
 \Bigl( {\g_-(-\a,-\b)\over\g_+(\a,-\b)}I + c.c. \Bigr),    \cr
}\eqn\IIz
$$
where
$ I=\int_{-\infty}^{\infty}dx\, g^\prime(x)\phipa_-(x)\phipa_+(x).$
Among these, the following unitarity relations hold:
$$\eqalign{
 T^s_{L\rightarrow L}+T^s_{L\rightarrow R}+R^s_{L\rightarrow R}=1, \qquad
 T^s_{R\rightarrow L}+T^s_{R\rightarrow R}+R^s_{R\rightarrow L}=1.
}\eqn\IIag
$$
\section{Fermion incident from broken-phase region and reciprocity${}^{5)}$}
In place of $\phipa_\pm(x)$, we start with $\phi^{(-\beta)}_\pm(x)$:
$$\eqalign{
\phi^{(-\beta)}_\pm(x)\rightarrow\e^{-\beta x},\qquad
\phi^{(+\beta)}_\pm(x)=\bigl(\phi^{(-\beta)}_\pm(x)\bigr)^*
\rightarrow\e^{+\beta x}
}\eqn\IIba
$$
at $x\rightarrow-\infty$, while
at $x\rightarrow+\infty$
$$\eqalign{
 &\phi^{(-\b)}_\pm(x)\sim {\widetilde \g}_\pm(-\b,\a)\e^{\a x}+{\widetilde
\g}_\pm(-\b,-\a)\e^{-\a x}, \cr
 &\phi^{(+\b)}_\pm(x)\sim {\widetilde \g}_\pm(\b,\a)\e^{\a x}+{\widetilde
\g}_\pm(\b,-\a)\e^{-\a x}.\cr
}\eqn\IIbb
$$
In parallel to Sect.2.3, we can derive the transmission and reflection
coefficients (denoted by the superscript $b$) such as
$T^b_{L\rightarrow L}={\bar T}^b_{R\rightarrow R}$ and
$R^b_{L\rightarrow L}=\bar R^b_{R\rightarrow R}=-R^b_{R\rightarrow R}=-\bar
R^b_{L\rightarrow L}$. The following unitarity relations hold as they should:
$$\eqalign{
 T^b_{L\rightarrow L}+T^b_{L\rightarrow R}+R^b_{L\rightarrow
L}+R^b_{L\rightarrow R}=1, \qquad
 T^b_{R\rightarrow L}+T^b_{R\rightarrow R}+R^b_{R\rightarrow
L}+R^b_{R\rightarrow R}=1.
}\eqn\IIbc
$$
\par
Since  $\phi^{(\pm \b)}_\pm(x)$ are linearly dependent on
$\phi^{(\pm \a)}_\pm(x)$, we have
${\widetilde \g}_\pm(\b,\a)=(\b/\a)\g_\pm(-\a,-\b)$ and so on. Then
we can prove reciprocity relations for the chiral fermion scattered off the
bubble wall including CP violation of $O(g^1)$:
$$\eqalign{
 T^b_{L\rightarrow R}+T^b_{R\rightarrow R}=
 T^s_{L\rightarrow L}+T^s_{L\rightarrow R}, \qquad
 T^b_{L\rightarrow L}+T^b_{R\rightarrow L}=
 T^s_{R\rightarrow L}+T^s_{R\rightarrow R}.
}\eqn\IIah
$$
\chapter{Quantum-Number Flux through Bubble Wall${}^{5)}$}
Let $Q_L$ ($Q_R$) be an additive quantum number carried by the 
left(right)-handed fermion, which is conserved in the symmetric
phase. The quantum-number flux into the symmetric-phase region just
in front of the bubble wall moving with velocity $u$ is given by
$$\eqalign{
F_Q= &{1 \over \gamma}\int_{m_0}^\infty dp_L \int^\infty_0 {dp_Tp_T
\over{4\pi^2}}
(Q_L-Q_R) \cr
     &\times \Bigl[(R^s_{R\rightarrow L}-{\bar R^s_{R\rightarrow
L}})f^s(p_L,p_T)-
(T^b_{L\rightarrow R}+T^b_{R\rightarrow R}-T^b_{L\rightarrow
L}-T^b_{R\rightarrow L})f^b(-p_L,p_T)\Bigr], \cr
}\eqn\IIIa
$$
where $\gamma=\sqrt{1-u^2}$ and the fermion-flux density in the symmetric
(broken) phase $f^s(f^b)$ is given by
$$\eqalign{
& f^s(p_L,p_T)=(p_L/E)\bigl(\exp {[\gamma (E-up_L)/T]}+1\bigr)^{-1}, \cr
& f^b(-p_L,p_T)=(p_L/E)\bigl(\exp {[\gamma (E+u{\sqrt
{p_L^2-m_0^2}})/T]}+1\bigr)^{-1}, \cr
}\eqn\IIIb
$$
the chemical potential being omitted for simplicity.
\par
Thanks to the unitarity and reciprocity relations, \IIIa\ is reduced to a
simple expression:
$$\eqalign{
F_Q= {1 \over \gamma}\int_{m_0}^\infty dp_L \int^\infty_0 {dp_Tp_T
\over{4\pi^2}}
(Q_L-Q_R)\Bigl[f^s(p_L,p_T)-f^b(-p_L,p_T)\Bigr]\Delta R.
}\eqn\IIIc
$$
Here $\Delta R$ is the difference between the chiral fermion and
its anti-fermion in the reflection coefficients incident from
the symmetric-phase region:
$$
\Delta R
\equiv R^s_{R\rightarrow L}-\bar R^s_{R\rightarrow L}
= 2R^{(0)s}(\delta^{inc}+\delta^{refl})
=-2T^{(0)s}\delta^{inc},
\eqn\IIah
$$
where
$$\eqalign{
 T^{(0)s}={\a\over\b}{1\over{\abs{\g_+(\a,\b)}^2}},  \qquad
 R^{(0)s}=\abs{\g_+(\a,-\b)\over\g_+(\a,\b)}^2
}\eqn\IIac
$$
with $T^{(0)s}+R^{(0)s}=1$ are respectively the transmission and reflection
coefficients in the absence of CP violation,
and we have used $\delta^{inc}+R^{(0)}\delta^{refl}=0$.
\par
We now recognize that $\Delta R$ is the dynamical quantity of primary
importance, since, if its absolute value be small, any quantum-number
flow is almost forbidden. Note that the vector-like quantum numbers
such as baryon and lepton numbers do not flow through the
CP-violating bubble wall. (The next order correction would be $O(g^3)$.)
Then one might have to resort to some charge transport scenario or other
to explain cosmological baryogenesis${}^{3,8)}$. If one takes hypercharge
flux, for example, $(Q_L-Q_R)=-1,1,1$ for the up-type quark, down-type
quark, massive lepton respectively.
\chapter{Quantum-Number Flux through the Kink-Type Bubble Wall}
A simple profile of the bubble wall without CP violation may be of the kink
type studied in Ref.6):
$$
f(x)=(1+\tanh x)/2.
\eqn\IIIIa
$$
The unperturbed solutions $\phi^{(\pm \a)}_\pm(x)$ are expressed in terms of
the hypergeometric functions. The transmission and reflection coefficients
without CP violation are respectively
$$\eqalign{
 &T^{(0)s}
={{\sin(\pi\a)\sin(\pi\b)}\over
  {\sin[{\pi\over2}(\a+\b+\xi)]\sin[{\pi\over2}(\a+\b-\xi)]}},\cr
 &R^{(0)s}
={{\sin[{\pi\over2}(\a-\b+\xi)]\sin[{\pi\over2}(\a-\b-\xi)]}\over
  {\sin[{\pi\over2}(\a+\b+\xi)]\sin[{\pi\over2}(\a+\b-\xi)]}}.\cr
}\eqn\IIIf
$$
\par
The effects of CP violation can be evaluated, once the functional form
of $g(x)$ is given. We have executed numerical calculations for several
forms of $g(x){}^{5)}$. We show one example in
Fig.1, {\foot {Because of a trivial error in our program for numerical
calculation, Figures in Ref.4) are incorrect.}} {\it i.e.,}
$g(x)=\Delta\theta\cdot f(x)^2$,
where $\Delta\theta$ characterizes the magnitude of CP violation.
(Note that $g(x)=\Delta\theta\cdot f(x)$ gives rise to no quantum-number
flow, as the $x$-independent phase is removed away.)
Although the sign of $\Delta R/\Delta\theta$ varies depending on functional
forms of $g(x)$, the dependence of
$\Delta R/\Delta\theta$ on the wall thickness $1/a$
shows a remarkable general trend. Namely,
${\abs{\Delta R/\Delta\theta}}$
is very small for $a/m_0 \lsim 0.5$, as typically illustrated in Fig.1.
Then it rapidly grows as $a$ increases but turns to decrease for $a/m_0
\gsim 2$.
\chapter{Concluding Remarks}
Although many complicated factors such as the wall velocity $u$ would
largely affect baryogenesis, a sufficient amount of baryon-number asymmetry
could not be produced unless ${\abs{\Delta R/\Delta\theta}}$
due to CP violation is large enough. We conclude with some remarks.
\par
(1) The general trend mentioned just before would require that, in order
for an effective baryogenesis to work, the wall thickness $1/a$ is smaller
than $2/m_0$, about twice the Compton wave length of the relevant fermion.
This would imply that, for $1/a=C/T$ with $C\simeq (20-40)$ of
one-Higgs-doublet models${}^{7)}$, any fermion whose mass $m_0$ is
subject to $m_0 \gsim 2T/C$ could not contribute to baryogenesis.
For $T\sim 200$ GeV, the top quark is excluded from the baryogenesis
game, even though its interaction with the bubble wall is large.
{\it A larger Yukawa coupling does not necessarily mean a larger effect
of CP violation.}
\par
(2) It is impossible for the one-Higgs-doublet model to incorporate
$x$-dependent phase into the bubble wall. On the other hand, there would be
many varieties of two-Higgs-doublet models or SUSY extension of the standard
model, among which a new source of CP violation $g(x)$ may be incorporated.
Once the wall thickness $1/a$ is given, it selects what species of fermions
take part in baryogenesis game. \par
(3) If some of the models predict a thin bubble wall, they may generally be
favored to explain the cosmological baryogenesis. We expect that our DWBA
prescription and the results would serve to build models to generate
baryon asymmetry of the universe.
\par
\bigskip
{\bf References}
\item{1)} A.~Sakharov, JETP Lett. {\bf 5} (1967) 24.
\item{2)} V.~Kuzmin, V.~Rubakov and M.~Shaposhnikov,
Phys.~Lett. {\bf B155} (1985) 36.
\item{3)} A.~Nelson, D.~Kaplan and A.~Cohen,
Nucl.~Phys. {\bf B373} (1992) 453;\nextline A.~Cohen,
D.~Kaplan and A.~Nelson,
Ann.~Rev.~Nucl.~Part.~Sci. {\bf 43} (1993) 27.
\item{4)} K.~Funakubo, A.~Kakuto, S.~Otsuki, K.~Takenaga and
F.~Toyoda, \ \ preprint\nextline
SAGA--HE--55---KYUSHU--HET--15 (1994),
to be published in Phys.~Rev.~{\bf D}.
\item{5)} K.~Funakubo, A.~Kakuto, S.~Otsuki, K.~Takenaga and F.~Toyoda,
\nextline in preparation.
\item{6)} A.~Ayala, J.~Jalilian-Marian, L.~McLerran and
A.~P.~Vischer, \ \ preprint \nextline
TPI--MINN--54---MUC--MINN--93/30--T---UMN--TH--1226/93 (1993).
\item{7)} M.~Dine, R.~Leith, P.~Huet, A.~Linde and D.~Linde,
Phys.~Rev.~{\bf D46} (1992) 550.\nextline
B-H.~Liu, L.~McLerran and N.~Turok, Pys.~Rev.~{\bf D46} (1992) 2668.
\item{8)} M.~Joyce, T.~Prokopec and N.~Turok, \ \ preprint
PUPT--91--1437 (1993).
\bigskip
{\bf Figure Caption}
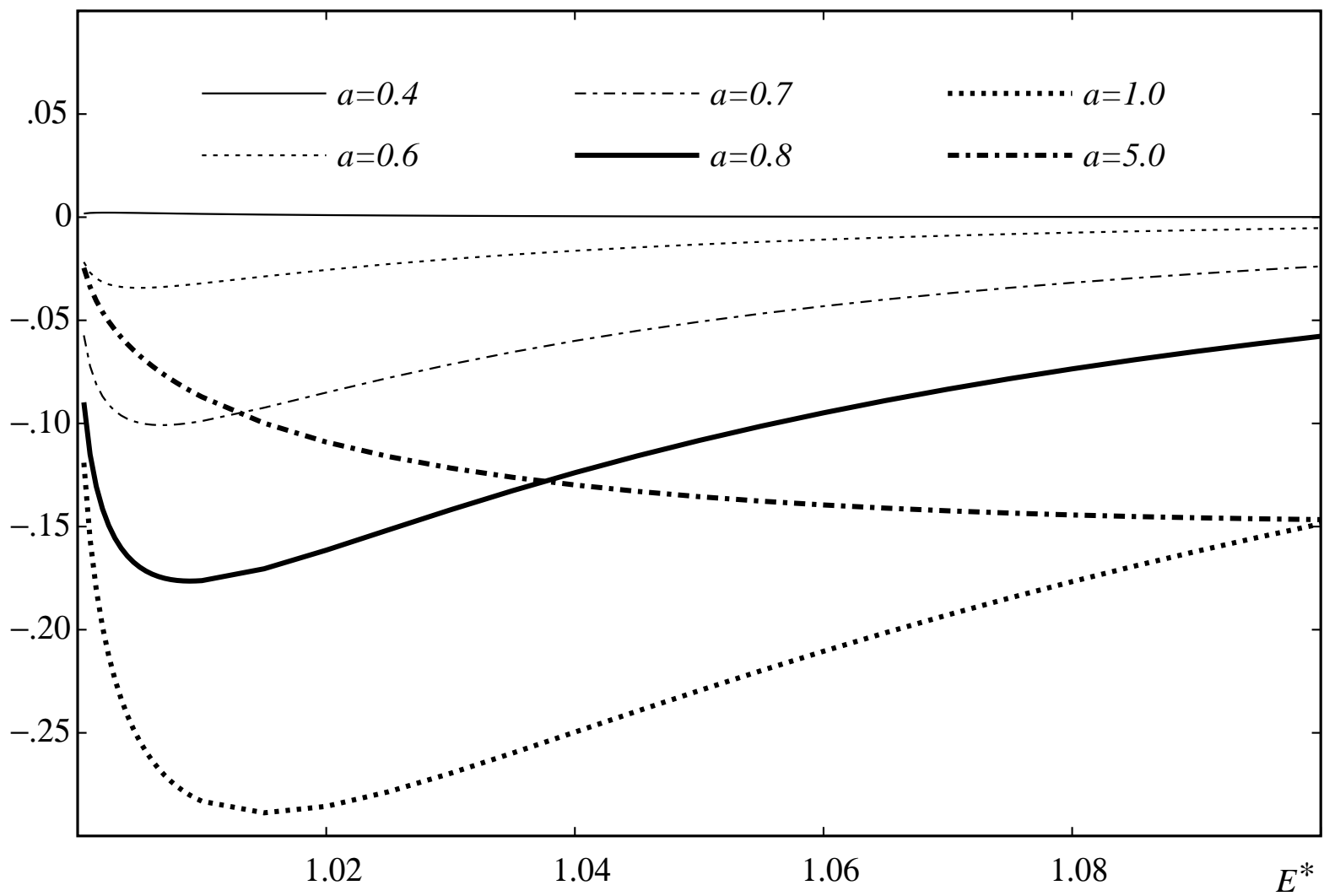
\item{\rm Fig.1} $\Delta R/\Delta \theta$ as a function of $E^*$ for
various $a$, in the case where $g(x) = \Delta \theta \cdot f(x)^2$.
The numerical values of $E^*$ and $a$ are respectively those of
$E^*/m_0$ and $a/m_0$.
\bye